\documentclass[aps,prl,floatfix,preprintnumbers,superscriptaddress,showkeys,showpacs,reprint,10pt]{revtex4-1}

\usepackage{graphicx}
\usepackage{fancyhdr}
\usepackage{amsmath,amssymb}
\usepackage{pdfpages}
\usepackage{aas_macros}
\usepackage{natbib}
 \usepackage[pdftex,
 				colorlinks=true,
 				citecolor=red,urlcolor=blue,
 				pdfpagelabels=true,
 				pdfstartview=Fit,
 				pdftitle={Systematic effects in the estimate of the local gamma-ray emissivity},
 				pdfauthor={Delahaye, Salati, Fiasson, \& Pohl},
				bookmarksopen=true,
				pdfpagelabels=true,
				]{hyperref}

\newcommand{\beq}{\begin{equation}}
\newcommand{\eeq}{\end{equation}}
\newcommand{\ben}{\begin{eqnarray}}
\newcommand{\een}{\end{eqnarray}}
\newcommand{\bi}{\begin{itemize}}
\newcommand{\ei}{\end{itemize}}

\begin{document}
\preprint{IFT-UAM/CSIC-11-62}

\title{Systematic effects in the estimate of the local gamma-ray emissivity}

\author{Timur Delahaye}
\email{timur.delahaye@uam.es}
\affiliation{Instituto de F\'isica Te\'orica UAM/CSIC
Universidad Aut\'onoma de Madrid
Cantoblanco, 28049 Madrid | Spain
}

\author{Pierre Salati}
\affiliation{LAPTH, Universit\'e de Savoie, CNRS, B.P.110 74941 Annecy-le-Vieux | France}

\author{Armand Fiasson}
\affiliation{LAPP, Universit\'e de Savoie, CNRS, BP110, F-74941 Annecy-le-Vieux Cedex | France}

\author{Martin Pohl}
\affiliation{Institut f\"ur Physik und Astronomie, Universit\"at Potsdam, Karl-Liebknecht-Strasse 24/25, 14476 Potsdam | Germany}
\affiliation{DESY, Platanenallee 6, 15738 Zeuthen | Germany}

\date{\today}

\begin{abstract}
We show in this letter that estimates of the local emissivity of $\gamma$-rays in the GeV-TeV range suffer uncertainties which are of the same order of magnitude as the current Fermi results. Primary cosmic-ray fluxes, cosmic-ray propagation, interstellar helium abundance and $\gamma$-ray production cross-sections all affect the estimate of this quantity. 
We also show that the so-called nuclear enhancement factor -- though widely used so far to model the $\gamma$-ray emissivity -- is no longer a relevant quantity given the latest measurements of the primary cosmic ray proton and helium spectra.
\end{abstract}
%

\maketitle


\section{Introduction}

The study of the Galactic $\gamma$-ray diffuse emission is of utmost importance as this component needs to be substracted to identify other extended astrophysical sources, such as the extra-galactic background or a possible dark matter component, and also because it is an interesting probe of the cosmic-ray population not only locally,
but all over the Galaxy.
In \citet{2011A&A...531A..37D}, we have shown how the morphology of the main component of the $\gamma$-ray diffuse emission (namely the one due to $\pi^0$ decay) is affected by the various uncertainties due to cosmic-ray propagation, gas distribution etc. Here we propose to look more closely into the specific features of the spectrum of this emission, especially in the light of the recent experimental data \citep{2009ApJ...703.1249A}.

The local $\gamma$-ray emissivity from $\pi^0$ decay has been thoroughly studied by the Fermi collaboration~\citep{2009ApJ...703.1249A} which found that the data are consistent with a high nuclear enhancement factor $\epsilon_M$ of 1.84 as found by~\citet{2009APh....31..341M} when considering the cross-section given by~\citet{2006ApJ...647..692K}. However we have already seen (see ref~\cite{2011A&A...531A..37D}) that these data seem to be in tension with the more recent cross-sections by~\citet{2007APh....27..429H}.

In this letter we investigate explanations of this discrepency as well as the various uncertainties affecting the estimate of the $\gamma$-ray emissivity. We quantify the impact of the primary cosmic-ray spectra, of their propagation, of the metallicity of the Interstellar medium (ISM) and of the production cross-sections.
For this, we follow the method developed in \citet{2011A&A...531A..37D}. As a reference case, we will consider the primary fluxes by \citet{2007APh....28..154S}, the med propagation parameters,  a helium to hydrogen ratio of 1/9, and the cross-sections by \citet{2006ApJ...647..692K} with the nuclear weights from \citet{Norbury:2006hp}. 
After having defined the various quantities at stake, we vary each parameter one by one to estimate its impact on the $\gamma$-ray emissivity and finally conclude.

\begin{figure}
\centering
\includegraphics[width=\columnwidth]{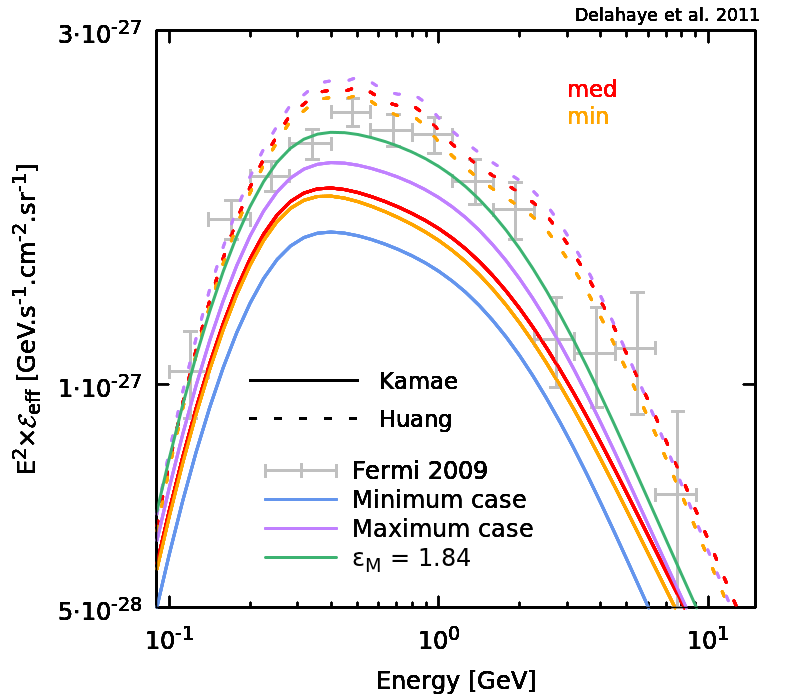}
\caption{\label{fig:un}
Effective emissivity as a function of energy. Plain lines correspond to \citet{2006ApJ...647..692K} cross-sections and dashed lines to~\citet{2007APh....27..429H} ones. Red and orange lines are respectively for med and min propagation parameters. The green line corresponds to the \citet{2006ApJ...647..692K} cross-sections used with a nuclear enhancement factor of 1.84 as suggested in the Fermi paper. The blue and purple lines correspond respectively to a minimum and a maximum case which are detailed in the text. Gray points are the recent Fermi data~\cite{2009ApJ...703.1249A}
}
\end{figure}

\section{Definition of the quantities of interest}

The hadronic $\gamma$-ray emissivity of an interstellar nucleus $A$
impacted by the cosmic-ray species $b$ is given by the convolution:
\beq
{\cal E}_{\rm A}^{b}(\mathbf{x},E)\!\! =\!\!
{\displaystyle \int}_{\!\!\! T_{\! min}}^{+ \infty} dT
\left\{
{\displaystyle \frac{d{\sigma}}{dE}}
\left(
b[T] + {\rm A} \to \gamma[E]
\right) \times \Phi_{b}(\mathbf{x},T) 
\right\} \; , \nonumber
\label{def:loc_emissivity_H}
\eeq
where $\Phi_{b}(\mathbf{x},T)$ is the flux of cosmic ray species $b$.
Because the $\gamma$-ray flux is correlated to the hydrogen number
density $n_{\rm H}$, it is more convenient to use the effective emissivity
per hydrogen atom defined as.
\beq
{\cal E}_{\rm eff}(\mathbf{x},E) \!\! = \!\!
{\displaystyle \sum_{\rm A}} \frac{n_{\rm A}}{n_{\rm H}} \;
\left\{
{\cal E}_{\rm A}^{p}(\mathbf{x},E) + {\cal E}_{\rm A}^{\alpha}(\mathbf{x},E)
\right\}. \nonumber
\eeq

The Fermi collaboration has selected two intermediate latitude regions in which the hydrogen is atomic and mostly local. The point sources and the inverse Compton component have been substracted substracted from the maps. The residual flux has been correlated to the HI column density, yielding an effective emissivity ${\cal E}_{\rm eff}(E)$ which could
be considered a priori equal to the solar value
${\cal E}_{\odot}(E ) \equiv {\cal E}_{\rm eff}(\mathbf{x}_{\odot},E )$. How local is the Fermi measurement is at the center of our analysis.
What Fermi actually measures is in fact quite close to the average of the
$\gamma$-ray emissivity ${\cal E}_{\rm eff}$ over the lines of sight of the
pixels of the maps.
What Fermi actually measures is in fact quite close to the average of the
$\gamma$-ray emissivity ${\cal E}_{\rm eff}$ over the line of sight. That is
why we have computed the emissivity inside each pixel by weighing it 
with the HI spatial density and averaged the results over the sky region investigated in the Fermi analysis.

\begin{figure}
\centering
\includegraphics[width=\columnwidth]{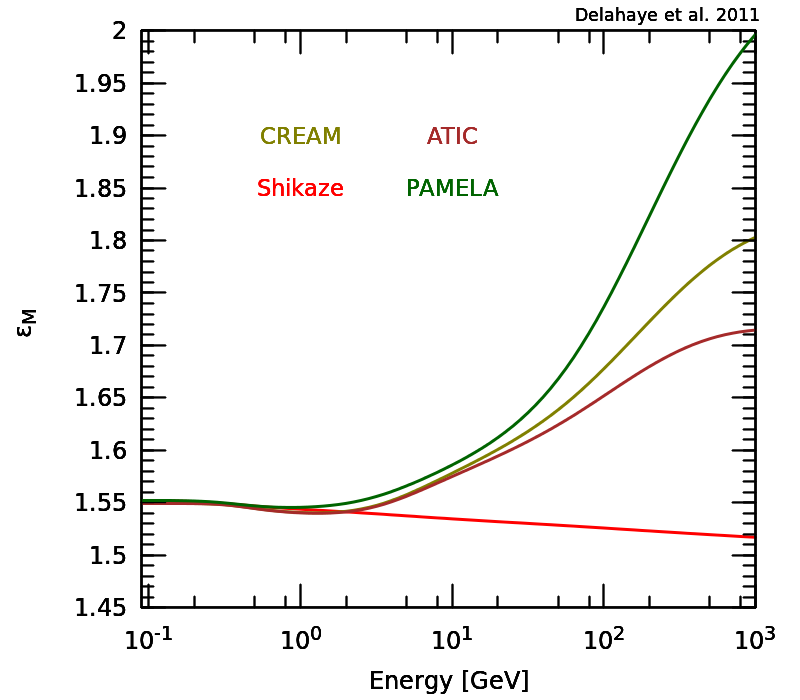}
\caption{\label{fig:trois}
Nuclear enhancement factor ploted as a function of the energy in the specific case of~\citet{2006ApJ...647..692K} cross-sections with nuclear weights from \citet{Norbury:2006hp}. Other models give similar results. The colored curves correspond to primary cosmic ray fluxes measured by BESS~\cite{2007APh....28..154S}, CREAM~\cite{2011ApJ...728..122Y}, ATIC~\cite{2009BRASP..73..564P} and Pamela~\cite{2011Sci...332...69A}.
}
\end{figure}

A quantity commonly used in the literature is the nuclear enhancement factor
which is defined as the ratio
$\epsilon_{M} = {\cal E}_{\rm eff}(\mathbf{x},E) /  {\cal E}_{\rm H}^{p}(\mathbf{x},E)$. The variations of $\epsilon_{M}$ with position $\mathbf{x}$ and energy $E$ have been so far disregarded and the nuclear enhancement factor has been mainly introduced as a constant by which proton-hydrogen interactions have to be renormalized in order to yield the total $\gamma$-ray flux. Even
at the Sun, it depends on the energy since it can be expressed as
\beq
\epsilon_{M}(E) \, = \, {\displaystyle \sum_{\rm A}} \;
{\displaystyle \frac{n_{\rm A}}{n_{\rm H}}} \; \left\{
w(1,{\rm A}) \; + \; w(4,{\rm A}) \,
{\displaystyle \frac{\Phi_{\alpha}(\odot,E)}{\Phi_{p}(\odot,E)}}
\right\} \;\; ,\nonumber
\label{simple_epsilon_M}
\eeq
where, $w($A$_1$,A$_2)$ are the nuclear weights by which the proton plus proton cross-section is multiplied to get the one of heavier elements collision. Untill recently, it was thought that the ratio of $\alpha$ to proton cosmic ray fluxes is constant with energy. However, recent results from \citep{2011ApJ...728..122Y,2009BRASP..73..564P,2011Sci...332...69A} indicate that this ratio is increasing at high rigidity. The exact explaination of this increase is still under investigation (see for instance~\cite{2011arXiv1105.4521B}) but has quite an impact on all secondary cosmic ray \citep{2011PhRvD..83b3014D,2011MNRAS.414..985L} fluxes. As one can see from Fig.~\ref{fig:trois}, $\epsilon_M$ may vary by more than 30\% from 1~GeV to 1~TeV making this quantity absolutely useless especially in the case where the fit to the PAMELA data proposed in the \hyperref[app]{appendix} are used.


\section{Propagation}

\begin{figure}
\centering
\includegraphics[width=\columnwidth]{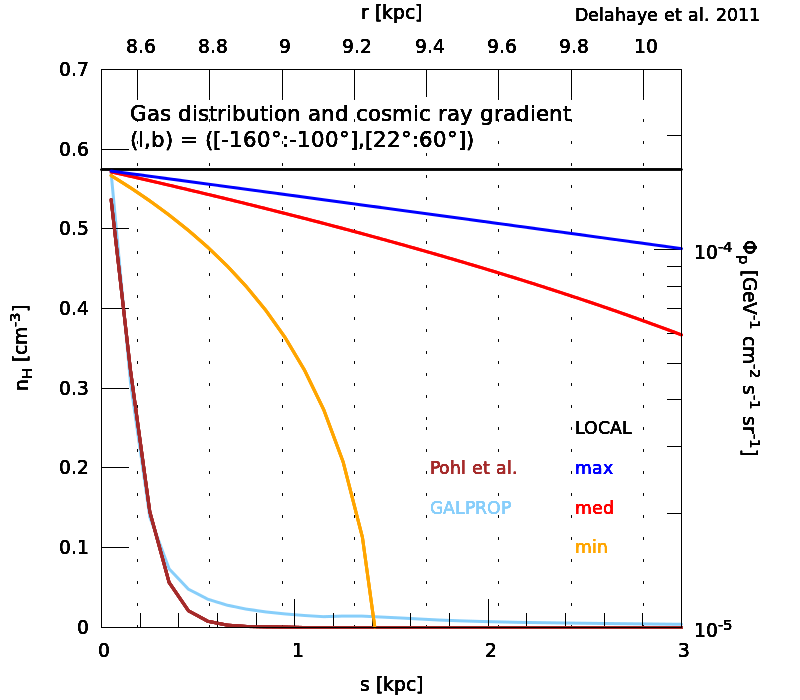}
\caption{\label{fig:six}
Gas density (left ordinate) and cosmic ray gradient (right ordinate) as a function of distance from the Sun (lower abscissa) or from the galactic center (upper abscissa).
}
\end{figure}

\begin{figure}
\centering
\includegraphics[width=\columnwidth]{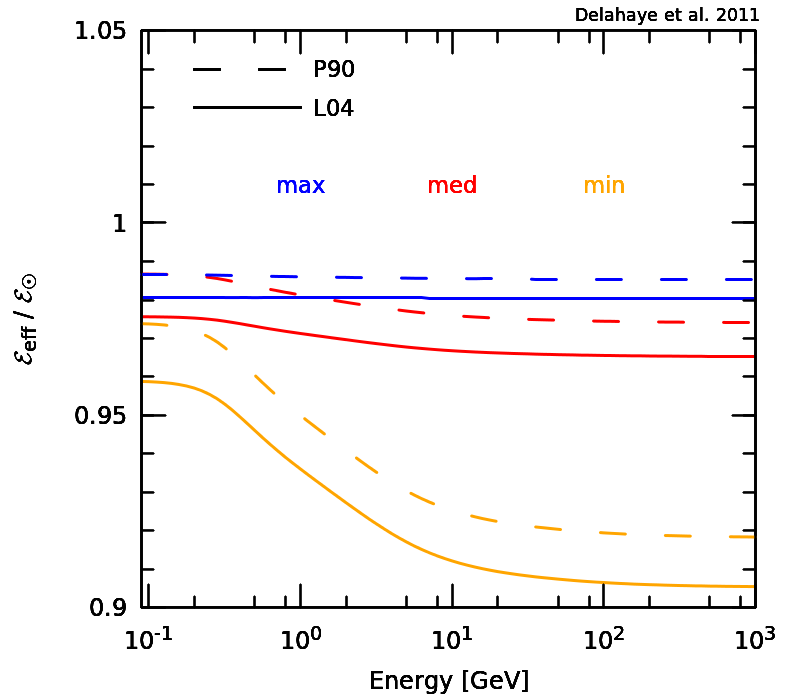}
\caption{\label{fig:deux}
Effective emissivity divided by solar emissivity as a function of energy. Plain lines correspond to source distributions by~\citet{2004IAUS..218..105L} and dashed lines to the one from~\citet{1990ApJ...348..485P}. Red, orange, and blue lines respectively correspond to med, min, and max propagation parameters sets.
}
\end{figure}

As one can see from Fig.~\ref{fig:six}, in the direction $l\in\left[-160^\circ;-100^\circ\right], b\in\left[22^\circ;60^\circ\right]$, which is the direction of interest of the Fermi study, the gas is mainly within 1~kpc from the Sun. However, depending on the propagation parameters chosen, the cosmic ray flux may exhibit a gradiant within this 1~kpc. This translates in the fact that ${\cal E}_{eff}$ can differ from ${\cal E}_{\odot}$ by up to 10\% especially at energies higher that 10~GeV (see Fig.~\ref{fig:deux}). 

In the present case, we have taken into account the cosmic-ray propagation within the framework developed in \citet{2011A&A...531A..37D} making use of the propagation parameters which give a good agreement with all cosmic-ray flux measurements~\cite{Maurin:2001sj,2010A&A...516A..66P}. Within this consistent framework three propagation parameter sets have been singled out~\cite{2004PhRvD..69f3501D}, they give a relatively good estimate of a median and two extreme cases labeled min, med and max in Figs.\ref{fig:un},\ref{fig:six} and \ref{fig:deux}. The med set is always used in this work unless specifically written otherwise.

Note however that the gradient of cosmic-ray in the direction of study not only depends on the propagation parameters but also on the cosmic-ray source profile. As one can see in Fig.~\ref{fig:deux} if we consider the less steep source profile by \citet{1990ApJ...348..485P} rather than the one by \citet{2004IAUS..218..105L}, the effect of propagation is quite reduced.

\section{Metallicity}
\label{sec:met}

\begin{figure}
\centering
\includegraphics[width=0.48\textwidth]{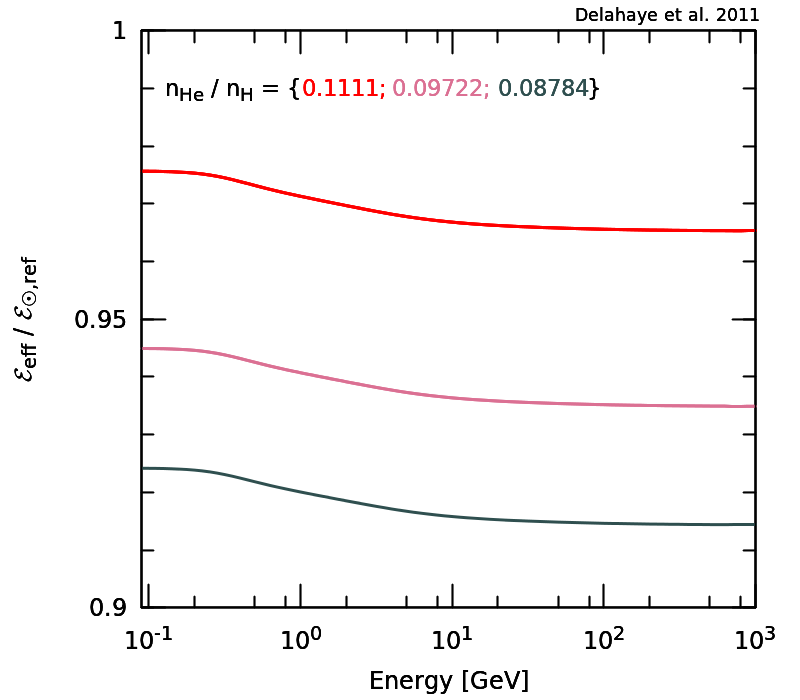}
\caption{\label{fig:cinq}
Effective emissivity divided by solar emissivity as a function of energy. Various values of the helium to hydrogen ratios for the interstellar gas. In red the value considered as a reference in this paper, in pink a value consistent with observation of the local region and in dark green a value consistent with observations of the anti-center. 
}
\end{figure}

The chemical composition of the interstellar medium, and what is more interesting in our case, the relative abundance of helium with respect to hydrogen is not precisely known. The cosmological value is about 0.077 He atom per H \cite{1997ApJ...483..788O} however, due to stellar evolution, the metallicity has evolved in the Milky Way. Because the stars do not have the same mass and age everywhere in the Galaxy, one should expect to have a helium gradient decreasing from the Galactic Center towards the outside regions. This is indeed supported by various observations~\cite{2000MNRAS.311..317C,2000MNRAS.311..329D} and theoretical models~\cite{2008RMxAA..44..341C}. On average one can consider that $n_{{\rm He}}/n_{{\rm H}}$ varies from $\sim$0.111 at the Galactic Center to $\sim$0.087 in the outer region with a local value of $\sim$0.097. These three values are illustrated in Fig.~\ref{fig:cinq}. The reason why we have considered a high metallicity as a reference model is only because it was the value taken by \citet{2007APh....27..429H}, allowing easier comparisons.

The $p+$He process and the $\alpha+$He one amount respectively for $\sim$20\% and $\sim$5\% of the total signal hence varying $n_{{\rm He}}/n_{{\rm H}}$ by 20\% translates in a variation of the total signal of 5\%, as one can see from Fig.\ref{fig:cinq}. 

\section{Cross-sections}
As a reference case (hereafter called KNT) we considered the cross-sections by \citet{2006ApJ...647..692K} (based on \texttt{P\textsc{ythia}} 6~\citep{2001hep.ph....8264S}) with the nuclear weights from \citet{Norbury:2006hp}. \citet{2007APh....27..429H} (but also \citet{2009APh....31..341M}, which lead to the value of $\epsilon_M=1.86$) is based on \texttt{DPMJET3}~\citep{2001ICRC....2..439R}. Both these Monte Carlo generators were not tuned with TeVatron and LHC data at the time of these works hence it is probable that one may expect some change in the future. Though \citet{2011arXiv1106.5073C} have shown that taking into account more recent versions of \texttt{P\textsc{ythia}} does not change much the results.

\begin{figure}
\centering
\includegraphics[width=\columnwidth]{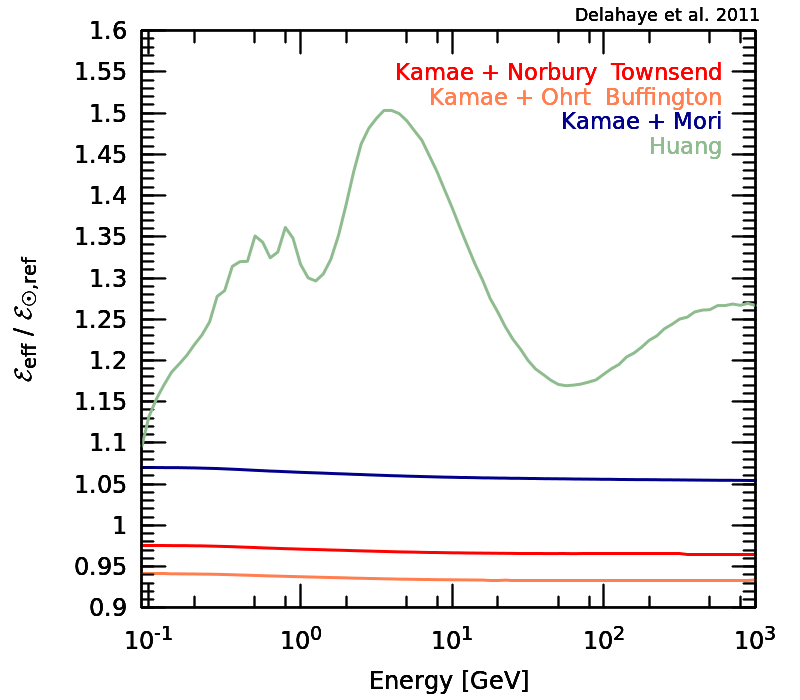}
\caption{\label{fig:quatre}
Effective emissivity divided by the reference solar emissivity (KNT) changing cross-sections and nuclear weights.
}
\end{figure}

However, considering what is available for the moment, as one can see in Fig.~\ref{fig:quatre}, the different models of nuclear weights and cross-sections have a large impact on the estimate of the $\gamma$-ray emissivity. In the case of the production cross-sections of \citet{2007APh....27..429H}, the difference reaches 50\%. Note that the tables available concern cross-sections of cosmic-ray protons and $\alpha$ which interact with a mixture of hydrogen and helium but also carbon, nitrogen and oxygen. It is hence not possible to change these proportion and it is hard to tell whether the differences come from the interstellar medium metallicity or from the Monte-Carlo.

The sharp features of the green line of Fig.~\ref{fig:quatre} are due to the resonnaces that \citet{2007APh....27..429H} have added to their cross-sections.

\section{Conclusion}

The various effects we have pointed out in this letter: primary cosmic-ray fluxes, propagation, and metallicity which lead to uncertainties of order 45\%, 10\% and 5\% respectively are illustrated in Fig.~\ref{fig:un} in the KNT case by the blue and purple lines which show the two extreme cases. On top of that, we have also considered the uncertainties due to $\gamma$-ray production cross-sections and nuclear weights which vary from 10\% to 50\%, depending on the energy. Hence the total theoretical uncertainty on the $\gamma$-ray production by $\pi^0$ decay is far from negligible.

The statistical uncertainties of the Fermi results vary from $\sim$4\% to $\sim$30\% which is of the same order than all the effects we have discussed here. Moreover, the Fermi data stop at 10~GeV whereas most models sensibly differ from each other at higher energies. It will hence be of great interest to have more data in the coming years. However, improving the experimental precision calls theoreticians to produce as much an effort to reduce the theoretical uncertainties we have pointed here.
 

\begin{acknowledgments}
We are deeply thankfull to Elisabeth Vangioni for her insight of the $^4$He Galactic gradient.
This work was supported by the Spanish MICINN’s  Consolider-Ingenio 2010
Programme under grant CPAN CSD2007-00042. We also acknowledge the support of the MICINN under grant FPA2009-08958, the Community of Madrid under grant
HEPHACOS S2009/ESP-1473, and the European Union under the Marie Curie-ITN program PITN-GA-2009-237920.
\end{acknowledgments}

\bibliography{local_em}

\appendix*
\section{Fit to the Pamela data}
\label{app}
In order to do the computations we mainly used the fit performed by~\cite{2007APh....28..154S} over the BESS data but also the fit done by Lavalle~\citep{2011MNRAS.414..985L} over data from ATIC~\citep{2009BRASP..73..564P} and CREAM~\citep{2010ApJ...714L..89A} balloons. Moreover, we propose here a fit to the recent PAMELA data~\cite{2011Sci...332...69A}: after demodulating in the force field approximation with a Fisk potential of 500MV:
\ben
\Phi_p(T) =&  35.3\, 10^{-4} \left(1 - \exp\left(-\left(\frac{T}{2.5 \mathrm{GeV}}\right)^{0.9}\right)\right) \left( \frac{T}{10 \textrm{GeV}}\right)^{-2.5} \nonumber\\ 
\times & \left( 1 + \frac{T}{16 \textrm{GeV}}\right)^{-0.5}  \left( 1 + \frac{T}{300 \textrm{GeV}}\right)^{0.46} \left( 1 + \frac{T}{5 \textrm{TeV}}\right)^{-0.21},\nonumber
\een
and
\ben
\Phi_\alpha(T) =&  1.5\times 10^{-5} \left( \frac{R}{50 \textrm{GV}}\right)^{-2.7} \left( 1 + \frac{R}{250 \textrm{GV}}\right)^{-1.3} \nonumber\\
\times & \left( 1 + \frac{R}{1 \textrm{TV}}\right)^{5.4} \left( 1 + \frac{R}{2 \textrm{TV}}\right)^{-4.15} ,\nonumber
\een
where $R$ stands for the rigidity of the particle and both fluxes are expressed in $(\textrm{(GeV/n).s.sr.cm}^2)^{-1}$.
\end{document}